# PAGE: a modern measure of emotion perception for teamwork and management research

Ben Weidmann & Yixian Xu*

*This paper presents a new measure of emotional perceptiveness called PAGE (Perceiving AI Generated Emotions). The test includes a broad range of emotions, expressed by ethnically diverse faces, spanning a wide range of ages. We created stimuli with Generative AI, demonstrating the potential to build customizable assessments of emotional intelligence at relatively low cost. Study 1 describes the validation of the image set and test construction. Study 2 reports the psychometric properties of the test. Despite its brevity (8 minutes on average) PAGE has strong convergent validity and moderately higher internal consistency than comparable measures. Study 3 explores predictive validity using a lab experiment in which we causally identify the contributions managers make to teams. PAGE scores strongly predict managers' causal contributions to group success, a finding which is robust to controlling for personality and demographic characteristics. We also discussed the potential of Generative AI to automate development of non-cognitive skill assessments.*

*Harvard Kennedy School. Corresponding author email: benweidmann@hks.harvard.edu. We would like to gratefully acknowledge the financial support of Schmidt Futures.

**Introduction**

The ability to recognize emotions is linked to a wide range of positive outcomes, including income (Momm et al., 2015), job performance (Côté, 2014; Elfenbein, 2023), leadership emergence (Côté et al., 2010), teamwork effectiveness (Farh et al., 2012; Weidmann & Deming, 2021) and negotiation (Elfenbein et al., 2007; Schlegel et al., 2018).

For over 50 years researchers have been developing skill-based tests to measure emotional perceptiveness. These tests generally ask participants to assess emotional expressions that are portrayed by actors in videos, images or audio recordings. The assessments have proved valuable in a wide range of disciplines including psychology, economics and medicine. However, existing measures suffer from three challenges that have limited their usefulness (summarized in Table 1). First, most tests use ethnically homogenous stimuli and often *only* include Caucasian faces. This can result in biased tests, as participants recognize emotions more quickly and accurately when the person expressing the emotions shares their cultural and ethnic identity (Elfenbein & Ambady, 2003; Young et al., 2012). Ethnically homogenous test stimuli are especially problematic when assessments are used in diverse populations. Second, many assessments lack emotional range and include only a handful of basic emotions – often including only one positive emotion. This constrains researchers' ability to assess emotion recognition ability in real world, in which people encounter a wide range of complex emotions (Kleef, 2016; Phillips & Slessor, 2011). Third, many existing tests have practical limitations that make them difficult for researchers to use – especially in field settings. These limitations include the length and cost of tests, along with the lack of freedom to use assessments on whatever platform researchers find convenient.

To address these limitations our paper develops a practical, freely available test that includes a wide range of emotions expressed by racially diverse faces, spanning ages 20-60. We create our stimuli with Generative AI, demonstrating the potential for building customizable assessments of emotional intelligence at relatively low cost. Leading models, such as OpenAI's DALL-E and Google's Imagen have advanced to a point where they can generate photorealistic images using simple text prompts on researcher-friendly platforms (Ramesh et al., 2022; Saharia et al., 2022). While much existing research focuses on the risks associated with the use of AI-generated content – for example disinformation campaigns, and AI-induced distrust (Jakesch et al., 2023; Miller et al., 2023; Nightingale & Farid, 2022), our study demonstrates a potential benefit: democratizing the ability to generate research stimuli to evaluate human capabilities. We use this low-cost tool to develop our inclusive emotion recognition assessment, called 'Perceiving AI Generated Emotions (PAGE)'.

**Table 1: Existing measures of emotional perceptiveness and their potential limitations**

| Test | Emotional Range | Ethnic Diversity | Practical Challenges | Item # |
|---|---|---|---|---|
| DANVA-2 (Nowicki & Duke, 1994) | 4 emotions | Caucasian, Black | Not freely available | 48 |
| BLERT (Bell et al., 1997) | 7 emotions | Caucasian | 15 – 20 minutes | 21 |
| JACBART (Matsumoto et al., 2000) | 7 emotions | Asian, Caucasian | Not freely available | 56 |
| RMET (Baron-Cohen et al., 2001) | 26 mental states | Caucasian | None | 36 |
| PERT-96 (Kohler et al., 2003) | 5 emotions | Diverse | None | 96 |
| MSCEIT Perception Tests (Mayer et al., 2003) | 5 emotions | Caucasian | Not freely available | 50 |
| MERT (Bänziger et al., 2009) | 10 emotions | Caucasian | 45 - 60 minutes | 120 |
| MiniPONS (Bänziger et al., 2011) | 2 affective situations | Caucasian | 15 – 20 minutes | 64 |
| ERI (Scherer & Scherer, 2011) | 5 emotions | Caucasian | 15 - 20 minutes | 60 |
| GERT-S (Schlegel & Scherer, 2016) | 14 emotions | Caucasian | 15 - 20 minutes; No customization | 42 |

Notes: DANVA: Diagnostic Assessment of Non-Verbal Abilities; BLERT: Bell Lysaker Emotion Recognition Task; JACBART: Japanese and Caucasian Brief Affect Recognition Test; RMET: The Reading the Mind in the Eyes Test; PERT-96: Penn Emotion Recognition Task; MSCEIT: Mayer–Salovey–Caruso Emotional Intelligence Test; MERT: Multimodal Emotion Recognition Test; MiniPONS: Profile of Nonverbal Sensitivity (short version); ERI: Emotion Recognition Index; GERT-S: Geneva Emotion Recognition Test (short version)

The rest of the paper proceeds as follows. Study 1 describes the construction of the test. Studies 2a and 2b assess its psychometric properties and examine convergent validity by reporting on the correlation between 'Perceiving AI Generated Emotions (PAGE)' and 'Reading the Mind in the Eyes Test (RMET)', a widely-used measure of emotion perception and theory of mind. Study 3 explores the predictive validity of the PAGE test. Using repeated random assignment of managers to groups, we examine the extent to which PAGE predicts the causal contribution that managers make to team success.

**Study 1**

Study 1 reports on the methods used to generate and validate the faces used in the PAGE instrument. We also describe other aspects of the test design including selection of distractors for the multiple-choice task and the scoring method.

**Methods**

***Generating images.*** Images were generated using DALL-E 2, a diffusion-based generative model that allows users to create photorealistic images from text prompts (Marcus et al., 2022; Ramesh et al., 2022). We chose DALL-E 2 because diffusion-based models typically outperform other generative models such as GANs in facial image generation (Stypułkowski et al., 2023).

***Emotion selection.*** Many emotion perception tasks only include the "basic 6 emotions" – anger, disgust, fear, happiness, sadness, and surprise (Ekman et al., 1969). However, recent research suggests that people can reliably recognize up to 28 emotions from facial-bodily expressions (Cowen & Keltner, 2020). For practicality and simplicity we focus on facial expressions. As such, we only include emotions that are less dependent on bodily expression and contextual clues. The 25 emotions we chose to experiment on generating with DALL-E 2 are six basic emotions, and 19 complex emotions: Disappointment, Amusement, Anxiety, Awe, Boredom, Concentration, Confusion, Contemplation, Contempt, Contentment, Desire, Doubt, Embarrassment, interest, Pain, Pride, Relief, Shame, Sympathy.

***Emotion prompt engineering.*** The prompts we gave DALL-E 2 used three methods, derived from emotion elicitation strategies researchers have used when creating emotional stimuli using human actors. The first method is simply to instruct expressers to express a particular emotion (as used, for example, by Lundqvist et al., 1998; Tottenham et al., 2009), e.g. "a 22 year old Caucasian woman feeling very angry". The second method relies on the Directed Facial Action Task (Ekman, 2007), in which expressers are instructed to employ specific facial actions based on the emotion prototypes identified by Ekman and colleagues (Ekman, 1992).[1] For example, to express the emotion *pride* we prompted DALL-E 2 by saying "a 30 year old Asian man showing pride. His head is held high, jaw thrust out, he has a small smile, lip pressed". Finally, we borrow a technique from studies of cultural variation in emotional expressions and use a short story to elicit emotions in expressers (Cordaro et al., 2018). For example, to generate an image of surprise, the prompt includes the following text "a 47 year old Indian woman showing a surprised face when hearing a loud sound she didn't expect".

We experimented with a combination of these three methods to elicit emotions in AI images, operationalized as prompts in three formats: '*emotion word*', '*emotion word, and facial actions*' '*emotion word, and one-sentence emotion story*.' To generate human-like images, each prompt begins with 'Generate a photorealistic image of…'. We also added 'detailed skin texture', and 'proportional eyes' into the prompt, which are found to be among the key factors in making an AI face look more realistic (Miller et al., 2023). An example prompt that uses all three of the techniques listed above is as follows.

---

[1] For examples of how this technique is used in generating images of faces with human actors, see Ekman and Friesens' Pictures of Facial Affect (Ekman & Friesen, 1976), and the Radboud Faces Database (Langner et al., 2010).

> "A realistic photo of a 20 year old Indian woman caught *embarrassed* and blushing in a social gaffe. Her whole face and head are in the middle. Plain grey background (leave some blank space around). She is wearing a white t-shirt. No body language, head oriented at the front, and staring at the camera."

A full list of prompts is provided for each emotion in Table 1, Appendix I.

***Stimuli standardization.*** We generated over 150 realistic faces for the initial stim set. These faces represent 25 emotions, six ethnicities,[2] and ages ranging from 20 to 60. We used Adobe Photoshop to standardize the stimuli to have consistent grey background. Images were resized so that each target's face and head are in the middle. See Figure 1 for sample stimuli.

**Figure 1: Example stimuli from PAGE**

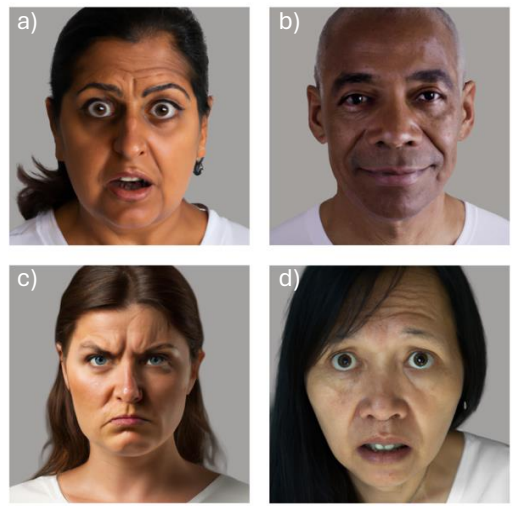

Note: The target emotions are: a) surprise, b) contentment, c) anger, d) fear

**Validation and Selection of Stimuli**
***Validating DALLE-2's emotional expressions using human raters.*** We recruited 500 participants on Prolific[3] to rate the stimuli in emotion categories. Each participant rated 30 - 35 images. Participants were asked to select one emotion that best described the face, from a list of 25 emotions. Each image was rated by at least 100 participants. Our sample was ethnically diverse and displayed gender balanced (see Table 3 for full demographic statistics). We computed the proportion of participants selecting the target emotion of each stimuli (out of 25 possible options) and retained those where a clear plurality of people identified the target emotion. From this sample, we selected a subset of 35 images, with the goal of producing a test that included racially diverse stimuli, gender equity and 20 emotions (see Table 2 for demographic characteristics of the images).

---
[2] We use the same ethnicity categories as the Chicago Face Database: Asian, Black, Caucasian, Indian, Latino, Multi-racial
[3] Douglas et al. (2023) compare several measures of data quality such as sample representativeness, and find that Prolific is comparable or better than competitors such as MTurk.

**Table 2: PAGE image count by demographics**

| Ethnicity | |
|---|---|
| Caucasian | 11 |
| Black | 8 |
| Latino | 9 |
| Asian | 4 |
| Indian | 2 |
| Multi-racial | 1 |
| Total | 35 |

| Age | |
|---|---|
| 20-29 | 5 |
| 30-39 | 16 |
| 40-59 | 13 |
| 60 | 1 |
| Total | 35 |

| Gender | |
|---|---|
| Female | 17 |
| Male | 18 |
| Total | 35 |

**Test Construction**

We generated a multiple-choice question for each image by selecting five distractors. Our distractors are emotions that are reliably recognizable in facial expressions (Cowen & Keltner, 2020). Emotions which are frequent parts of social interactions such as *confusion*, *doubt*, and *interest* (Benitez-Quiroz et al., 2016; Rozin & Cohen, 2003) are overrepresented in the distractors. This design choice reflects a desire to aid the predictive validity of the PAGE in real-world settings – especially those requiring teamwork.

The resulting set of 35 test questions were sequenced such that consecutive items did not feature the same emotion. Correct answers are scored as 1, incorrect as 0. All materials are freely available. We also administered a short and a full version of PAGE task on the website of Harvard Skills Lab[4] for easy public access. See Figure 2 for one example item of PAGE viewed from the lab website.

---

[4] PAGE (short version): https://game.skillslab.dev/experiments/1a9bf1fe-f6ac-46a7-a37a-c2c921be5d79; PAGE (full version): https://game.skillslab.dev/experiments/c5a9f233-1802-421d-8498-3a22228a057d. The short version was created by ranking items by retaining one item for each emotion, and selecting the item that ranked most highly in terms of the item's correlation with the score on the long test.

**Figure 2: Example item from PAGE**

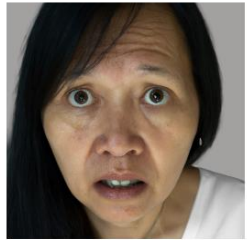

Note: example item from the PAGE test, with correct answer 'Fear'

**Study 2a: Measurement properties of Perceiving AI Generated Emotions (PAGE)**

We recruited 1010 participants from Prolific. All participants were located in the United States and were ethnically diverse (see Table 3 for details). We oversampled non-Caucasian participants so that we could better assess performance on the PAGE of people from different ethnic backgrounds. We also focused on respondents who were full-time workers (91%) aged 25-55, to validate the PAGE among a sample that would be of use to labor economists and organizational psychologists. We administered the PAGE test on Qualtrics (for full instructions, see Appendix II) and each participant received a compensation of $2.50. To motivate participants to maintain attention throughout the task, we also awarded the quintile of performers a $5 bonus. The median participant spent 8 minutes on the test.

**Table 3: Demographics of participants**

|  | **Study 1** | **Study 2a** | **Study 2b^** | **Study 3*** |
|---|---|---|---|---|
| **Number** | 500 | 1010 | 741 | 116 |
| **Ethnicity (%)** | | | | |
| White | 57.8 | 44.5 | 50.5 | 17.0 |
| Black/African American | 5.8 | 22.3 | 21.5 | 16.1 |
| Latino/Hispanic° | 3.4 | 15.4 | 13.2 | - |
| Asian° | 20.0 | 17.9 | 14.8 | 56.2 |
| Other / not reported | 0.0 | 0.0 | 0.0 | 10.7 |
| **Age** | | | | |
| Mean (SD) | 34.0 (9.4) | 36.7 (9.1) | 37.6 (9.1) | 25.4 (4.5) |
| 18-29 (%) | 41.8 | 26.4 | 23.6 | 83.6 |
| 30-39 (%) | 32.6 | 36.6 | 35.2 | 16.4 |
| 40-59 (%) | 25.2 | 36.9 | 41.2 | - |
| 60-74 (%) | 0.4 | - | - | - |
| **Female (%)** | 49 | 50 | 50 | 43 |
| **Full-time workers (%)** | 46 | 91 | 100 | - |
| **Country** | US | US | US | UK |

Notes: ^Study 2b is a subset of study 2a. *This is the sample of managers in Study 3. The full experiment contained 555 participants. °This includes 'Asian British'; [0]In Study 3, which was done in the UK, this was not an option.

The mean score for PAGE is 23.7 (*SD* = 5.0). There was no evidence of ceiling or floor effects, as shown in Figure 3. Women (M = 24.0, *SD* = 5.1) performed slightly better than men (mean difference = 0.6, p = 0.06). There was a negative correlation between PAGE scores and age (r = -0.14, p < 0.001). These findings are consistent with previous work showing on-average female advantage in emotion recognition (Greenberg et al., 2023) and lower accuracy at emotional recognition in older adults (Mill et al., 2009; Ruffman et al., 2008).

**Figure 3: Distribution of scores by gender**

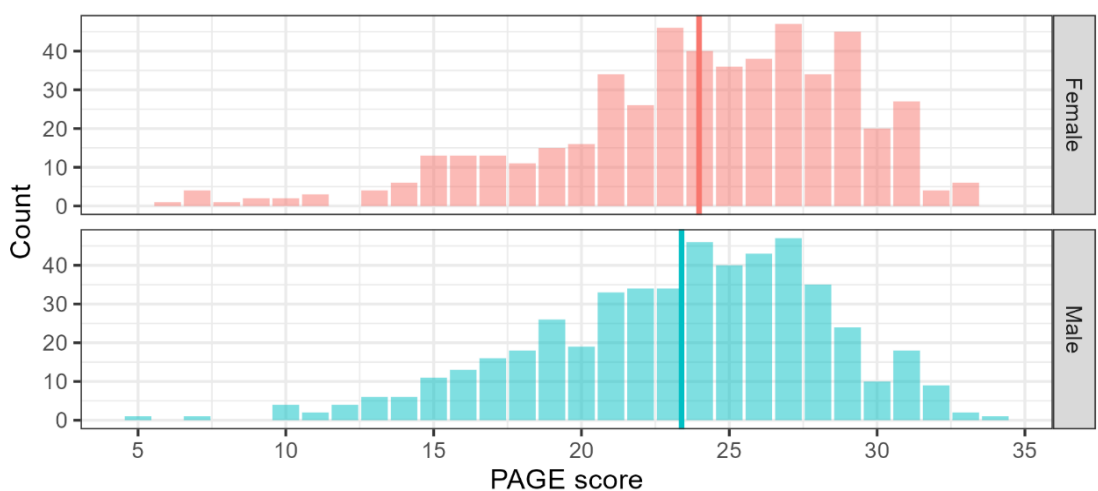

Cronbach's alpha = 0.73 for the PAGE test. This is substantially better than the average reported reliability of other emotion recognition ability (ERA) tests of $\alpha$ = 0.60 (Boone & Schlegel, 2016). We consider the internal consistency of PAGE to be a strength, particularly given the brevity of the test (8 minutes on average) and wide range of emotions included in the test.

Exploratory factor analysis suggests that the PAGE has a one-factor structure, see Figure 1 for the scree plot (Appendix I). To further assess the unidimensionality of the PAGE instrument, we tested a one-factor confirmatory factor analysis (CFA) model. Model fit was evaluated by inspecting the comparative fit index (CFI), the Tucker–Lewis Index (TLI), and the root mean square error of approximation (RMSEA). The one-factor CFA model showed good fit ($\chi^2$ = 884, $df$ = 560, $p$ = .000, CFI = .829, TLI = .818, RMSEA = .024). Although the CFI and TLI are slightly lower than common acceptability threshold (0.9), the low RMSEA and satisfactory values of Cronbach's alpha overall suggest that the PAGE is a unidimensional test. The factor structure of PAGE is consistent with findings about the dimensionality of emotion recognition abilities, that ERA is one broad ability consisting of correlated valence-based skills and minor ability facets related to pairs of similar and highly confused emotions (Hall, 2001; Schlegel et al., 2012).

**Study 2b**

In Study 2b, we examined the convergent validity of PAGE, by estimating the correlation of PAGE scores with Reading the Mind in the Eyes Test. This test is very similar to the PAGE in that it is a 36-item multiple choice test measuring emotional perceptiveness. The images in the RMET are faces which are cropped, to only include the eyes. The RMET is used across a range of disciplines, has been cited over 7,000 times, and was identified as the best existing measure of Theory of Mind in the Social Cognition Psychometric Evaluation (SCOPE) survey (Pinkham et al., 2014).

***Procedures and results*** We analyze a sub-sample of 741 participants from Study 2a, who completed both PAGE and RMET (administered on Qualtrics). Participant demographics are presented in Table 3 (see 'Study 2a'). Both tasks included one practice question to familiarize participants with the task format. To limit the effect of differences in vocabulary, we provided a list of emotion definitions for reference. When participants put their cursor above the emotion word, they were provided with a definition. Question order was randomized for both tests.[5] On average, participants completed the PAGE in 8 minutes and the RMET in 10 minutes. We find that the PAGE is highly correlated with RMET (raw correlation = .66, disattenuated correlation = .88, p < .001), providing strong evidence of convergent validity.

**Study 3**

To assess the predictive validity of PAGE we fielded the test in a lab experiment that used a novel design to identify the causal contribution that individual managers make to group performance. In the experiment, managers are randomly assigned to multiple teams. Our design relies on the repeated random assignment of managers to teams to identify the average impact each manager has on group performance (following the procedure in Weidmann & Deming, 2021). A total of 116 managers in the experiment completed the PAGE instrument, which allowed us to compare PAGE scores with causally identified manager contributions. We find that PAGE scores are predictive of manager contributions, and that this association is robust to controls for personality and demographic factors.

**Procedures**

This section provides a high-level overview of the experiment (for details, see Weidmann et al. 2024). We recruited an ethnically diverse sample of graduate and undergraduate students at the University of Essex in the UK. The median participant was 25 years old and had 2 years work experience. See Table 3 (in Study 2a) for demographic details.

---

[5] To reduce the impact of order effects, we had 249 participants complete the RMET first (then the PAGE) and 492 people complete the tests in the reverse order. Reliability and factor structure was extremely similar in both cases.

The goal of the experiment was to causally identify the contribution that managers made to teams, and to explore the characteristics that were predictive of management performance. The experiment included both individual and group assessments (see Figure 4 for experiment overview). Individual tests included a demographic questionnaire and a personality assessment (a shortened version of the Big5 inventory, described in Gosling et al., 2003). A subsample of experimental participants (n=116 managers) completed the PAGE assessment as well as the Reading the Mind in the Eyes Test (RMET).

Group testing took place at Essex Lab. Groups were hierarchical and consisted of 1 'manager' and 2 'workers'. At the start of the experiment, participants were assigned to the role of manager or worker. These roles persisted throughout the experiment. After receiving a role, participants were randomly assigned to groups of 3 people. Each group consisted of a manager and two workers. Groups worked face-to-face on a novel collaborative problem-solving assessment called the 'Collaborative Production Task' (full details are available in Weidmann et al., 2024). In the task, groups solved problems across three modules: numerical, spatial, and analytical reasoning. Managers were responsible for assigning workers to different modules, monitoring group progress, and keeping their team engaged. The task takes around 15 minutes, including dedicated time for participants to introduce themselves and familiarize themselves with the task. After each group finished the task, managers were randomly assigned another set of two workers. Over the course of the experiment each manager was randomly assigned to four groups.

**Figure 4: Overview of study 3**

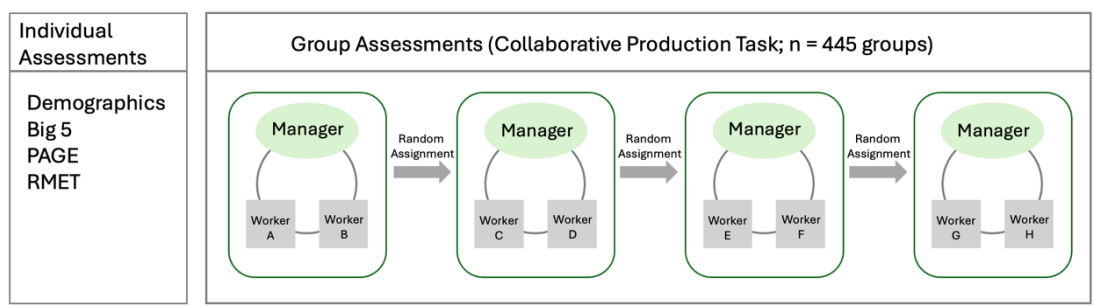

**Results**

First, we find that the PAGE score of each manager is positively associated with group performance. The correlation between manager PAGE scores and group performance = 0.16 (p<0.001, n=445 groups). These results are consistent with findings from non-hierarchical teamwork settings where team members' emotion perception ability positively predicts group performance (Elfenbein, Polzer, et al., 2007; Weidmann & Deming, 2021).

Next, we move from the group level to the level of individual managers. We identify the average causal impact each manager has by using the fact managers are randomly assigned to multiple teams. Because of this random assignment, we can estimate manager causal contributions by calculating the average performance of each manager's teams (Weidmann & Deming, 2021). We find that PAGE scores positively predict the average causal contribution that managers have on their groups (correlation = 0.28, p=0.002, n=116 managers). This association is robust to demographic and personality characteristics of managers (as indicated in Table 4).

We interpret the positive relationship between PAGE scores and manager contributions in light of the responsibilities that managers have in Study 3, which include monitoring and motivating their team. Managers who are better at perceiving emotions are more likely to notice when their teammates are confused and doubtful and require extra instruction about what to do. They may also be faster at perceiving low morale, and responding with support, motivation or switching the task a teammate is working on.

Last, we compare the ability of PAGE and RMET to predict manager contributions. The results are presented in Table 4. Overall, we find that the association between PAGE scores and manager contributions is greater than that for RMET, and that the association is robust to controls for differences in Big 5 personality measures and demographic factors (age, gender, ethnicity and education). Column 1 presents the raw association between manager causal contribution and PAGE scores, which are standardized to have mean=0 and sd=1. A 1sd increase in PAGE scores is associated with a shift in manager contributions of 0.29sd. Columns 2 and 3 add controls for Big 5 personality and demographics which reduces the coefficients slightly, but they remain large and significant. Columns 4 to 6 repeat this process, focusing on RMET as a predictor. The relationship between RMET and manager causal contributions are weaker and less robust. Column 7 is a full specification in which we include all variables, illustrating the robust association between PAGE scores and the impact that managers have on their teams. We speculate that two strengths of PAGE – the diversity of face stimuli and the focus on social emotions – make it a better predictor of manager contributions than RMET, especially for a participant sample that is predominantly non-Caucasian (72.3%) and a setting focused on interacting with people.

**Table 4: associations between manager performance and emotional perceptiveness**

| | Average Causal Contribution of Managers | | | | | | |
|---|---|---|---|---|---|---|---|
| | (1) | (2) | (3) | (4) | (5) | (6) | (7) |
| PAGE | 0.290*** | 0.242** | 0.231** | | | | 0.276** |
| | (0.094) | (0.098) | (0.110) | | | | (0.118) |
| RMET | | | | 0.146 | 0.071 | -0.027 | -0.123 |
| | | | | (0.089) | (0.092) | (0.111) | (0.116) |
| Big5 | | X | X | | X | X | X |
| Demographics | | | X | | | X | X |
| Observations | 116 | 112 | 110 | 116 | 112 | 110 | 110 |
| $R^2$ | 0.077 | 0.167 | 0.221 | 0.023 | 0.124 | 0.185 | 0.230 |
| Adjusted $R^2$ | 0.069 | 0.120 | 0.115 | 0.015 | 0.074 | 0.075 | 0.116 |

Notes: *p<0.1, **p<0.05, ***p<0.001. Dependent variable is each managers' estimated causal contribution, as measured by the average score across each manager's randomly assigned teams. Demographic factors include age, ethnicity, education and gender. PAGE and RMET scores are both standardized to have mean=0 and sd=1.

**Discussion and conclusion**

This paper develops and validates a measure of emotional perceptiveness using a demographically diverse set of 35 faces, expressing 20 emotions. Study 1 demonstrates that generative AI is capable of producing standardized, realistic faces that express both basic and complex emotions. Study 2a shows that the PAGE test is unidimensional and has relatively strong internal validity for a short test – especially when one considers both test length and the fact that PAGE covers an unusually wide set of emotions. Study 2b demonstrates convergent validity by showing that PAGE scores are very strongly associated with scores on RMET – a practical and widely used alternative measure of emotional perceptiveness. Finally, Study 3 provides evidence of predictive validity, especially for researchers interested in leadership and management, by showing that PAGE scores are strongly associated with the causal contribution that managers make to group performance in a controlled lab study. We find that PAGE is a better predictor of manager contributions than RMET. As RMET only contains Caucasian faces, the difference in performance may in part be due to the fact that the participant sample in Study 3 is predominantly non-Caucasian.

For these reasons, we believe that the PAGE test will be useful to researchers who need a reliable, short, skill-based measure of emotional perceptiveness that is appropriate for diverse populations. We view it as a useful general tool. We also hope that the PAGE instrument illustrates the potential for generative AI to help create targeted and customized measures of emotional intelligence by substantially reducing the cost of test creation and automating the test development process.

There are at least two ways in which tests may be usefully customized. First, researchers and practitioners may find it helpful to be able to vary the demographic profile of the stimuli. For example, tests of emotional intelligence fielded in school settings may want to focus on much younger faces, while those fielded in aged-care settings focus on old faces. Second, it may be beneficial to have tests that oversample specific complex emotions – many of which are absent from most measures. For example, an organization hiring a team leader may screen for the ability to recognize *confusion,* as this potentially enables quick clarification. Similarly, an ability to perceive *doubt* could help a manager identify the need for further persuasion or buy-in. In contexts where teamwork is important, perceiving *anxiety* may signal a colleague's need for support. Of course, it is an empirical question whether such customized measures of emotional recognition are more predictive of positive outcomes in real-world contexts, but with the advent of generative AI this research agenda is much more practically achievable.

Further improvements in AI technology are likely to make the process of creating tests of emotional perceptiveness faster and easier. In creating PAGE we hand-crafted multiple choice questions using the image stimuli generated by DALL-E. However, generative AI is already capable of creating multiple-choice questions across difficulty levels (Aryadoust et al., 2024). More significantly, large language models (LLMs) can potentially be utilized to simulate human participants and produce human-like results (Horton, 2023). If so, this allows test developers to combine a small sample of people with a large and very low-cost sample of LLM respondents to rigorously assess and refine the psychometric properties of new tests. Overall, while existing AI technology considerably reduced the practical barriers we faced in creating PAGE, it seems likely that these barriers will be progressively lowered.

In closing, we believe that the PAGE test measures a general construct that is an important determinant of success in a wide range of social activities, from negotiation and hiring, to networking and working in a team. The test has strong measurement properties, is appropriate for diverse populations, and is open access. We hope that others will build on the approach of using generative AI to create and validate customized tests that allow for a better understanding of the role emotion plays in facilitating interaction in the workplace and beyond.

# Appendix I

**Table 1: Emotion Prompts for PAGE Stimulus Generation**

| Emotion | Prompt | Method | Sources |
|---|---|---|---|
| Amusement | "Generate a photorealistic image of a [age] [ethnicity] [gender] laughing with jaw dropping, head tilting backwards, with detailed skin texture and natural lighting, with highly realistic, well-proportioned eyes, with opened eyes. No body language, showing the face and shoulder, head oriented at the front, and looking at the camera. Plain grey background, wearing a white t-shirt." | facial actions | Keltner, 1995 |
| Anger | "A realistic photo of a [age] [ethnicity] [gender] feeling very **angry**. Symmetric eyes. No body language, face in the middle, head oriented at the front, and staring at the camera. Plain grey background, wearing a white t-shirt" | emotion word | Ekman, 2007 |
| Anxiety | "Create a hyper-realistic image of a [age] [ethnicity] [gender] showing expression **anxiety**. Eyes looking sideways, frowned eyebrows, biting lips. Detailed skin texture and natural lighting. Wearing a white t-shirt. No body language, showing only the face, head oriented at the front, and staring at the camera. Plain grey background." | emotion word + facial actions | Perkins, Inchley-Mort, Pickering, Corr, & Burgess, 2012 |
| Boredom | "Create a hyper-realistic image of a [age] [ethnicity] [gender] showing expression **boredom**. Eyelids dropping. Detailed skin texture and natural lighting. Wearing a white t-shirt. No body language, showing only the face, head oriented at the front, and staring at the camera. Plain grey background." | emotion word + facial actions | Scherer & Ellgring, 2007 |
| Concentration | " A realistic photo of a [age] [ethnicity] [gender] feeling very **concentrated**, clearly paying attention to something intently. No body language, face in the middle, head oriented at the front, and staring at the camera. Plain grey background, wearing a white t-shirt No body language, face in the middle, head oriented at the front, . Plain grey background, wearing a white t-shirt" | emotion word + emotion story | Rozin & Cohen, 2003 |
| Confusion | "Create a hyper-realistic image of a [age] [ethnicity] [gender] showing a **confused** expression with slightly opened mouth. Detailed skin texture and natural lighting. No body language, showing only the face, head oriented at the front, and staring at the camera. Plain grey background, he is wearing a white t-shirt." | emotion word + facial actions | Rozin & Cohen, 2003 |
| Contemplation | "A realistic photo of a [age] [ethnicity] [gender] expressing the emotion **contemplation**, he is pondering life. No body language, face in the middle, head oriented at the front, and staring at the camera. Plain grey background, wearing a white t-shirt" | emotion word + emotion story | Rozin & Cohen, 2003 |
| Contempt | "Create a hyper-realistic photo of a [age] [ethnicity] [gender] expressing **contempt**. Detailed skin texture and natural lighting. No body language, showing | emotion word | Matsumoto & Ekman, 2004 |
| | only the face, head oriented at the front, and staring at the camera. Plain grey background, wearing a white T-shirt." | | |
| Contentment | "A realistic photo of a [age] [ethnicity] [gender] experiencing a feeling of **well-being and delight**. His whole face and head in the middle. Plain grey background (leave some blank space around). He is wearing a white t-shirt. No body language, head oriented at the front, and staring at the camera." | Synonym of emotion word | Cordaro, Brackett, Glass, & Anderson, 2016 |
| Disappointment | "Create a hyper-realistic image of a [age] [ethnicity] [gender] showing expression **disappointment**. Eyebrows slightly furrowed, lips pressed, eyes looking sideways. Detailed skin texture and natural lighting. Wearing a white t-shirt. No body language, showing only the face, head oriented at the front, and staring at the camera. Plain grey background." | emotion word + facial actions | Cordaro et al., 2016 |
| Disgust | "A realistic photo of a [age] [ethnicity] [gender] feeling **disgusted**. No body language, face in the middle, head oriented at the front, and staring at the camera. Plain grey background, wearing a white t-shirt" | emotion word | Ekman, 2007 |
| Doubt | "Create a hyper-realistic image of a [age] [ethnicity] [gender] showing a **doubtful** expression with pressed lips. Detailed skin texture and natural lighting. No body language, showing only the face, head oriented at the front, and staring at the camera. Plain grey background, he is wearing a white t-shirt." | emotion word + facial actions | Benitez-Quiroz et al., 2016 |
| Embarrassment | "A realistic photo of a [age] [ethnicity] [gender] caught **embarrassed** and blushing in a social gaffe. Her whole face and head in the middle. Plain grey background (leave some blank space around). She is wearing a white t-shirt. No body language, head oriented at the front, and staring at the camera." | emotion word + facial action + emotion story | Keltner, 1995 |
| Fear | "Create a hyper-realistic image of a [age] [ethnicity] [gender] showing expression **fear**. Wearing a white t-shirt. Detailed skin texture and natural lighting. No body language, showing only the face, head oriented at the front, and staring at the camera. Plain grey background, | emotion word | Ekman, 2007 |
| Interest | "Create a hyper-realistic image of a [age] [ethnicity] [gender] showing expression **interest**. His eyebrows pulled straight up, eyes open wide, he has a small smile, his head tilts forward,. Detailed skin texture and natural lighting. No body language, showing only the face, head oriented at the front, and staring at the camera. Plain grey background, he is wearing a white t-shirt." | emotion word + facial action | Reeve, 1993 |
| Joy | "A realistic photo of a [age] [ethnicity] [gender] expressing emotion **joy**, she is very happy at something unexpected. No body language, showing only the face, head oriented at the front, and staring at the camera. Plain grey background, wearing a white t-shirt"" | emotion word + emotion story | Ekman, 2007 |

| Pain | "Create a hyper-realistic image of a [age] [ethnicity] [gender] showing a **painful** expression. Her eyes closed tightly, her lips tighten and pressed. Detailed skin texture and natural lighting. No body language, showing only the face, head oriented at the front, and staring at the camera. Plain grey background, he is wearing a white t-shirt."" | emotion word + facial action | Prkachin, 1992 |
|---|---|---|---|
| Pride | "Create a hyper-realistic image of a [age] [ethnicity] [gender] showing **pride**. His head holds high, jaw thrusts out, he has a small smile, lip pressed. Detailed skin texture and natural lighting. No body language, showing only the face, head oriented at the front, and staring at the camera. Plain grey background, wearing a white t-shirt." | emotion word + facial action | Tracy & Robins, 2004 |
| Sadness | "A realistic photo of a [age] [ethnicity] [gender] showing a **sad** face when hearing an old friend's death. her whole face and head in the middle. Plain grey background (leave some blank space around). she is wearing a white t-shirt. No body language, face in the middle, head oriented at the front, and staring at the camera." | emotion word + emotion story | Ekman, 2007 |
| Surprise | "A realistic photo of a [age] [ethnicity] [gender] showing a **surprised** face when hearing something she didn't expect. her whole face and head in the middle. Plain grey background (leave some blank space around). she is wearing a white t-shirt. No body language, face in the middle, head oriented at the front, and staring at the camera." | emotion word + emotion story | Ekman, 2007 |

**Table 2: PAGE stimuli, target emotions, and distractors**

| # | Stimuli | Target emotion | Distractors in multiple-choice test |
|---|---|---|---|
| 1 | 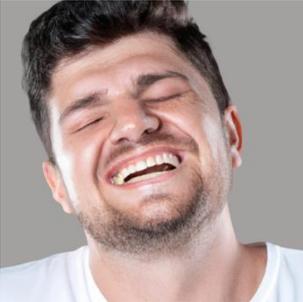 | Amusement | Awe, Pleasure, Interest, Surprise, Relief |
| 2 | 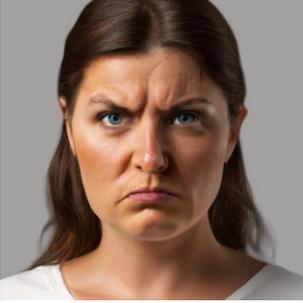 | Anger | Pride, Pain, Disgust, Confusion, Shame |
| 3 | 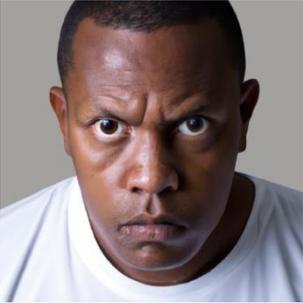 | Anger | Pride, Pain, Confusion, Sadness, Disgust |
| 4 | 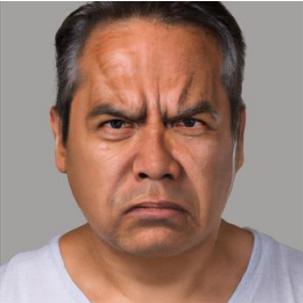 | Anger | Amusement, Contempt, Sadness, Disappointment, Doubt |

| 5 | 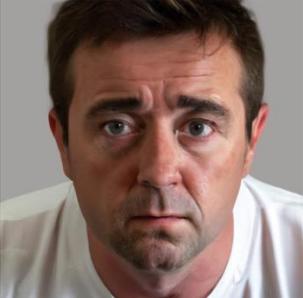 | Anxiety | Contentment, Embarrassment, Contemplation, Confusion, Fear |
| 6 | 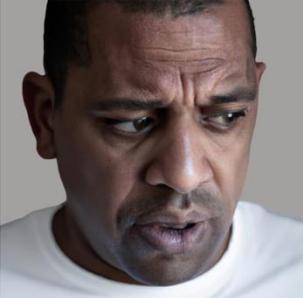 | Anxiety | Contentment, Disappointment, Disgust, Relief, Boredom |
| 7 | 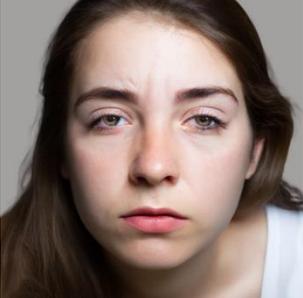 | Boredom | Interest, Distress, Pleasure, Pain, Anger |
| 8 | 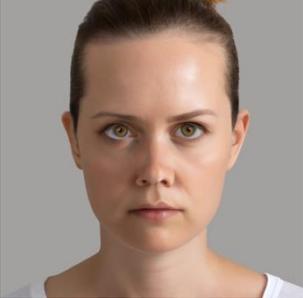 | Concentration | Confusion, Doubt, Contentment, Disappointment, Interest |
| 9 | 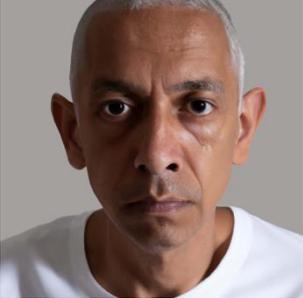 | Concentration | Contentment, Interest, Contempt, Anger, Disappointment |

| 10 | 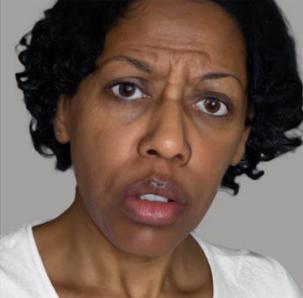 | Confusion | Surprise, Interest, Anxiety, Doubt, Disgust |
| 11 | 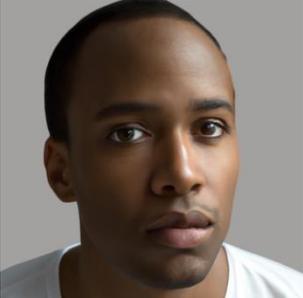 | Contemplation | Confusion, Surprise, Disappointment, Interest, Contentment |
| 12 | 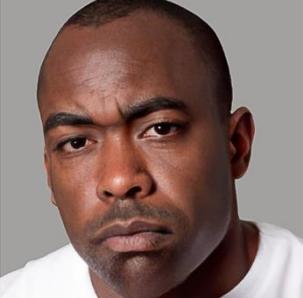 | Contemplation | Anxiety, Relief, Surprise, Interest, Contentment |
| 13 | 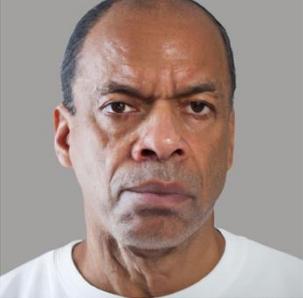 | Contempt | Anxiety, Disgust, Confusion, Interest, Boredom |
| 14 | 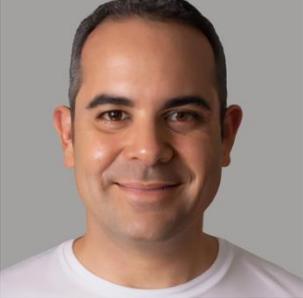 | Contentment | Ecstasy, Pride, Desire, Contemplation, Contempt |

| 15 | 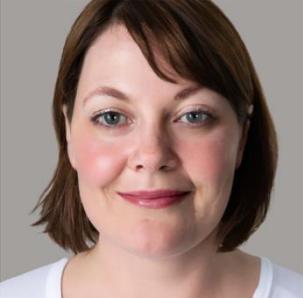 | Contentment | Ecstasy, Disappointment, Pride, Contempt, Relief |
| --- | --- | --- | --- |
| 16 | 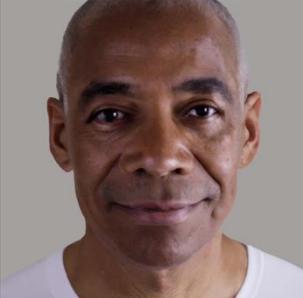 | Contentment | Amusement, Pride, Desire, Joy, Contempt |
| 17 | 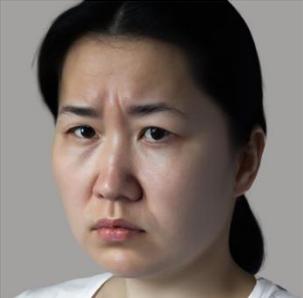 | Disappointment | Confusion, Anger, Disgust, Contempt, Boredom |
| 18 | 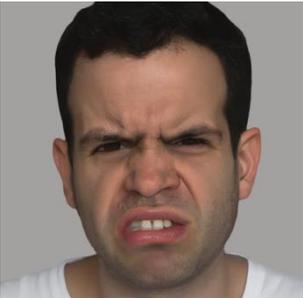 | Disgust | Amusement, Pain, Contempt, Confusion, Anger |
| 19 | 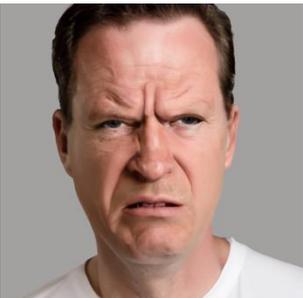 | Disgust | Contemplation, Pain, Contempt, Confusion, Anger |
| 20 | 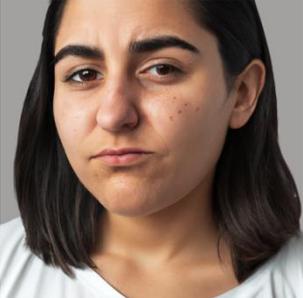 | Doubt | Contentment, Confusion, Anger, Contemplation, Interest |
| 21 | 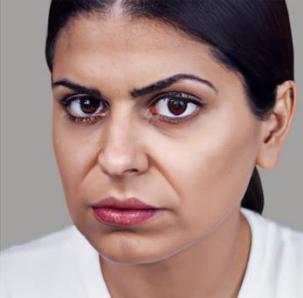 | Doubt | Interest, Confusion, Boredom, Sadness, Anxiety |
| 22 | 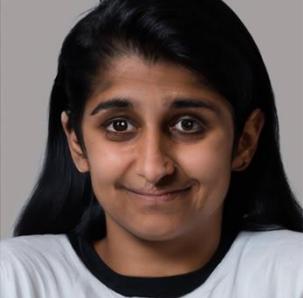 | Embarrassment | Relief, Confusion, Pride, Anxiety, Shame |
| 23 | 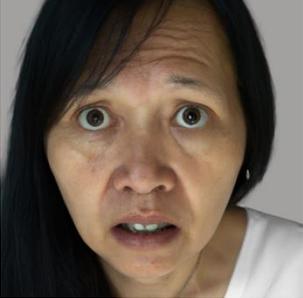 | Fear | Awe, Confusion, Surprise, Anxiety, Shame |
| 24 | 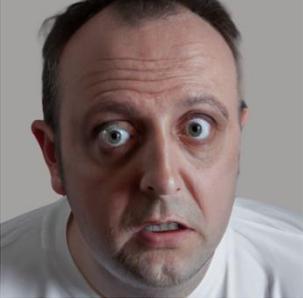 | Fear | Anger, Confusion, Awe, Embarrassment, Pleasure |

| 25 | 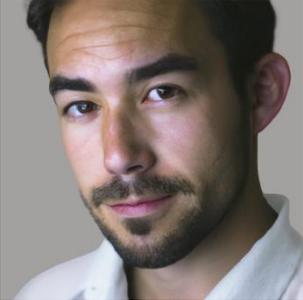 | Interest | Amusement, Boredom, Doubt, Contemplation, Sympathy |
| --- | --- | --- | --- |
| 26 | 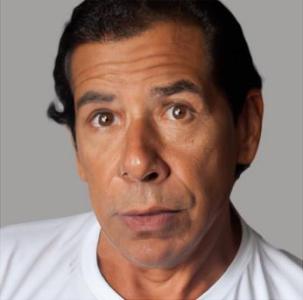 | Interest | Doubt, Boredom, Embarrassment, Surprise, Disappointment |
| 27 | 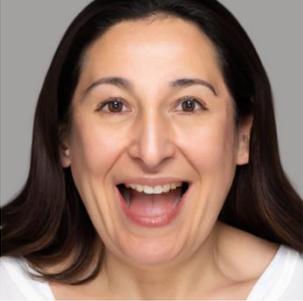 | Joy | Awe, Surprise, Desire, Contentment, Confusion |
| 28 | 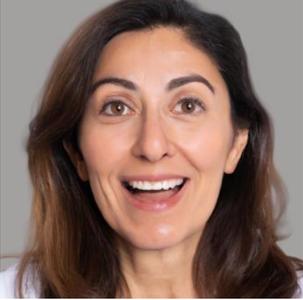 | Joy | Surprise, Contentment, Desire, Confusion, Awe |
| 29 | 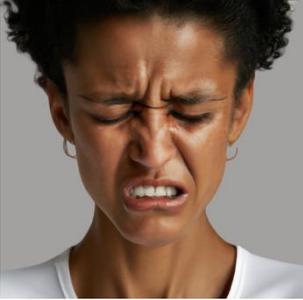 | Pain | Pride, Disappointment, Anger, Embarrassment, Shame |

| 30 | 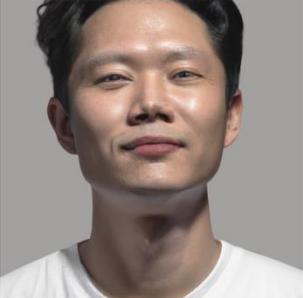 | Pride | Amusement, Awe, Interest, Joy, Contempt |
| --- | --- | --- | --- |
| 31 | 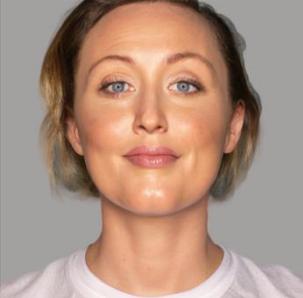 | Pride | Amusement, Doubt, Interest, Contentment, Contempt |
| 32 | 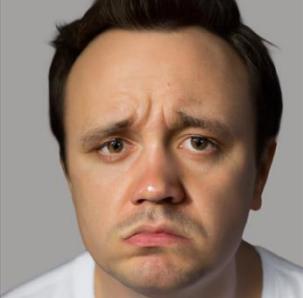 | Sadness | Contentment, Anxiety, Confusion, Pain, Boredom |
| 33 | 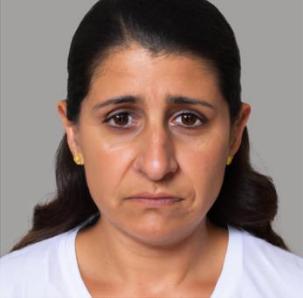 | Sadness | Contentment, Anxiety, Disappointment, Pain, Boredom |
| 34 | 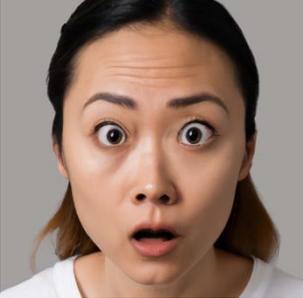 | Surprise | Interest, Ecstasy, Confusion, Anger, Fear |

| 35 | 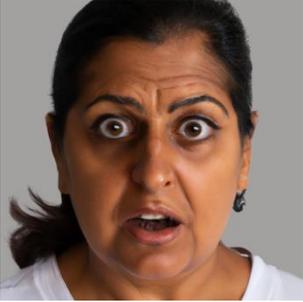 | Surprise | Interest, Anger, Relief, Confusion, Disgust |

**Figure 1: Scree plot of PAGE**

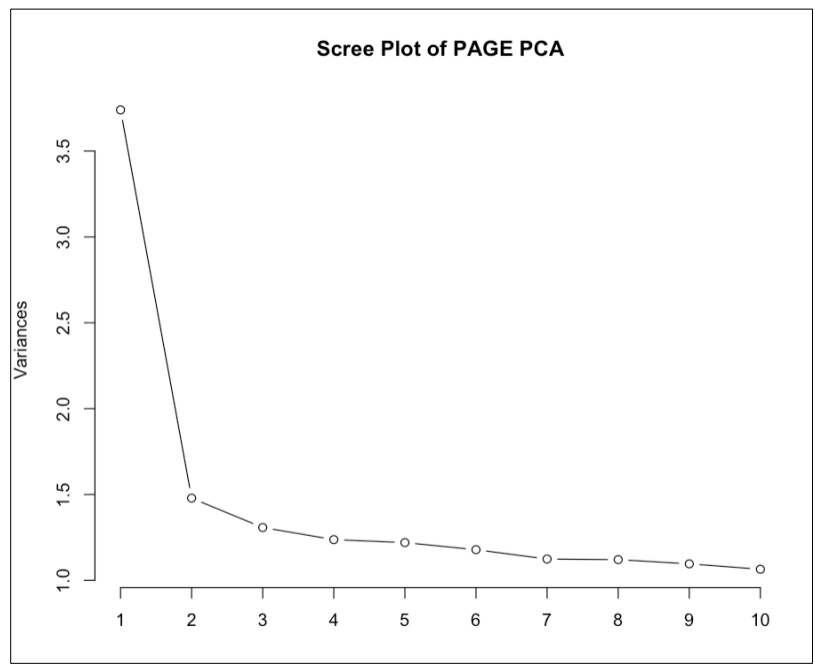

**Appendix II**

## PAGE task instructions

> Survey Completion 0% — 100%
>
> **Which emotion is being expressed?**
>
> In this task, you will see 35 facial images. Your goal is to accurately select the emotion which best describes the face.
>
> There may be instances when the emotion is not immediately clear. In such cases, please choose just one word, the word which you consider to be most prominently expressed on the face.
>
> Continue >>

Under each image, 6 emotion words are presented. Before making your choice, make sure that you have read all 6 words. To know the definition of each emotion word, hover your cursor over **the specific word** for seconds and its definition will appear. See the example screenshot.

You can also look it up in the list of definitions. Please open this link in a new tab to reference during the survey. Note that not all emotions listed are expressed on these faces.

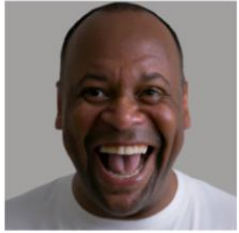

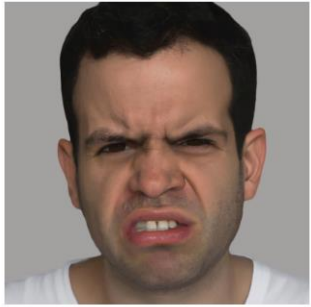

Please select the emotion that **best** describes the emotion on this face

○ Confusion  ○ Pain  ○ Anger
○ Disgust  ○ Contempt  ○ Amusement

Continue >>